\input phyzzx
\hsize=417pt 
\sequentialequations
\Pubnum={ EDO-EP-11}
\date={ \hfill April 1997}
\titlepage
\vskip 32pt
\title{ Cosmic Censorship in Quantum Gravity }
\vskip 12pt
\author{Ichiro Oda \footnote\dag {E-mail address: 
sjk13904@mgw.shijokyo.or.jp}}
\vskip 12pt
\address{ Edogawa University,                                
          474 Komaki, Nagareyama City,                        
          Chiba 270-01, JAPAN     }                          
%
%
%
%
%
\abstract{ We study cosmic censorship in the Reissner-Nordstrom charged 
black hole by means of quantum gravity holding near the apparent 
horizons. We consider a gedanken experiment whether or not a black hole 
with the electric charge $Q$ less than the mass $M$ ($Q < M$) could 
increase its charge to go beyond the extremal limit $Q = M$ by 
absorbing    
the external charged matters. If the black hole charge could exceed the 
extremal value, a naked singularity would be liberated from the 
protection of the outer horizon and visible to distant observers, which 
means weak cosmic censorship to  
be violated in this process. It is shown that the black hole never 
exceeds the extremal black hole this 
way in quantum gravity as in classical general gravity. An increment of 
the trapped external charged matters by the black hole  
certainly causes the inner Cauchy horizon to approach the outer horizon, 
but its relative approaching speed gradually slows down and stops 
precisely at the  
extremal limit. It is quite remarkable that cosmic censorship remains 
true even in quantum gravity. This study is the first attempt of examining 
weak cosmic censorship beyond the classical analysis.
 } 
\endpage
%
%
%

\def\sp(#1){\noalign{\vskip #1pt}}

%
%
%
%
%
\topskip 30pt
\par
\leftline{\bf 1. Introduction}	
\par
In recent years, we have had a recurrence of interests on a quantum 
theory of black holes. Many people have worked on the Hawking radiation 
and the fate of the endpoint of black hole evaporation [1], the 
information loss  
paradox [2], the black hole thermodynamics [3], in particular an 
understanding of the statistical-mechanical origin of black hole entropy 
e.t.c. by various different approaches.

Closely related to those problems, there exists one of the most important 
unsolved problems in classical general relativity, which is Penrose's 
cosmic censorship conjecture [4]. Penrose has advocated a stronger form 
of cosmic censorship, what nowadays we call, strong cosmic censorship, 
which insists roughly that the whole spacetime should be globally 
hyperbolic. Saying more precisely, the globally hyperbolic development 
$D(\Sigma)$ is inextendible for generic initial data on compact or 
asymptotically flat partial Cauchy surface $\Sigma$. In other words, the 
Cauchy horizon $\partial D(\Sigma) = 0$. In previous works, we have 
studied this strong cosmic censorship by evaluating the quantum 
generalized affine parameter (QGAP) and shown that the spacetime is 
effectively globally hyperbolic due to the violent quantum fluctuation of 
the geometry [5, 6].

There is a weaker version of cosmic censorship, what is called, weak 
cosmic censorship, which states in physical terms that singularities 
arising in generic gravitational collapse are hidden behind the event 
horizon. In this paper, when referring to cosmic censorship merely we 
always have weak cosmic censorship in mind. Of course, cosmic censorship 
conjecture is unsolved but is widely believed to be true for suitably 
arranged general initial data and equations of state. This is because 
without validity of cosmic censorship not only one cannot prove the 
no-hair theorem of a black hole [7] but also the fact that a black hole 
can have intrinsic entropy in itself becomes dubious [8].

An early attempt of examining cosmic censorship has been performed in the 
Reissner-Nordstrom charged black hole [9]. Let us imagine a physical 
situation where there is a static charged black hole which has the mass 
$M$ larger than the charge $Q$, that is, $M > Q$ in the geometrized 
units, for which we have two horizons, the inner Cauchy horizon and the 
outer event horizon. If one adds a stream of charged test particles with 
a large electric charge rather than its mass into this black hole, the 
initial black hole would turn to the black hole with the property of $M < 
Q$, where there is no horizon and a naked singularity is visible from 
external observers.  If so, this would lead to a clean counter-example to 
cosmic censorship conjecture. But it was shown that one needs a lot of 
kinetic energy to exceed the extremal limit $M = Q$ by this method so 
that the net increase of black hole mass is always larger than that of 
black hole charge [9]. Thus this implies that one cannot violate cosmic 
censorship in this physical setting. 

The main topic that we would like to address in 
this paper is that the above classical test of cosmic censorship remains 
true also in a quantum theory of general relativity. It is well known 
that while the outer horizon is a surface of infinite redshift, the inner 
Cauchy horizon is of infinite blueshift for our own asymptotically flat 
universe. Therefore as the infalling lightlike matters approach the inner 
Cauchy horizon the energy density of it will suffer an infinite 
blueshift, by which in the vicinity of the Cauchy horizon the curvature 
becomes extremely huge and quantum gravitational effects would play a 
dominant role. This is one of our motivations of attempting to analyze 
cosmic censorship by means of quantum gravity.

At present, as is well known, we do not have a good grasp of a fully 
satisfactory and mathematically consistent theory of quantum gravity yet. 
However, it has been recently pointed out that one can construct 
models of quantum black holes in the spherically symmetric geometry at 
least near the horizons and/or the singularities, and then the models 
were applied fruitfully to clarify physically interesting problems such 
as the Hawking radiation [10, 11, 12], the mass inflation [13] and the 
quantum instability of the black hole singularity in three dimensions 
[14]. The key observation behind these works is that the problems 
associated with quantum black holes have at all events an intimate 
relationship  
with the quantum-mechanical behavior of the horizons and/or the 
singularities and are largely irrelevant to the other regions of 
spacetime so that it might be sufficient to consider a quantum theory 
of black holes holding in their vicinities in order to answer the questions.

The article is organized as follows. In section 2, we review the 
canonical formalism of a spherically symmetric system with a black hole. 
This canonical formalism is used to construct the canonical formalism 
holding in the vicinity of the apparent horizons in section 3. In section 
4, based on the formalism built in section 3, we analyse the Hawking 
radiation from purely quantum-mechanical viewpoint. Here it is shown that 
the mass loss rate is proportional to $M^{-2}$ whose result is identical 
to the result inferred by Hawking [1]. Section 5 is the main part of this 
article where it is shown that for the Reissner-Nordstrom black hole, 
cosmic censorship never be violated also even in quantum gravity as in 
classical general relativity. The last section is devoted to 
conclusion.

\vskip 32pt
\leftline{\bf 2. Review of Canonical Formalism}	
\par
We start by reviewing a canonical formalism of a spherically symmetric 
system with a black hole. This canonical formalism was previously 
constructed by Hajicek et al. [15], and recently extended to the case of 
$\partial \over \partial r$ being timelike in the interior of a black 
hole by us [11]. Our attention in this article lies in the spacetime 
regions where the Killing vector field $\partial \over \partial x^0$ is 
timelike, thus we foliate the spacetime geometry by a family of 
spacelike hypersurfaces $x^0 = const$ according to the conventional 
procedure [16, 17]. 

The action that we consider is of the form
$$ \eqalign{ \sp(2.0)
S = \int  d^4 x \sqrt{-^{(4)}g} \ \bigl[ {1 \over 16 \pi G} {}^{(4)}R - 
{1 \over  
4 \pi} {}^{(4)}g^{\mu\nu} (D_{\mu} \Phi)^{\dag} D_{\nu} \Phi -  {1 \over 
16 \pi } F_{\mu\nu}F^{\mu\nu} \bigr],
\cr
\sp(3.0)} \eqno(1)$$
where $\Phi$ is a complex scalar field with the covariant derivative  
$$ \eqalign{ \sp(2.0)
D_{\mu} \Phi = \partial_{\mu}\Phi + i e A_{\mu} \Phi,
\cr
\sp(3.0)} \eqno(2)$$
$F_{\mu\nu}$ is $U(1)$ field strength as usual given by
$$ \eqalign{ \sp(2.0)
F_{\mu\nu} = \partial_{\mu} A_{\nu} - \partial_{\nu} A_{\mu},
\cr
\sp(3.0)} \eqno(3)$$
and $e$ is a positive electric charge of $\Phi$. To show the four 
dimensional  
character explicitly we append the suffix $(4)$ in front of the metric 
tensor and the  
curvature scalar. We follow the conventions adopted in the MTW textbook 
[18] and use the natural units $G = \hbar = c = 1$. The Greek indices 
$\mu, \nu, ...$ take the values 0, 1, 2, and 3, on the other hand, the 
Latin indices  $a, b, ...$ take the values 0 and 1. 
Of course, the inclusion of other matter fields, the cosmological 
constants and the surface terms in this formalism is straightforward even 
if we limit ourselves to the action (1) for simplicity.

Let us take the following spherically symmetric assumptions
$$ \eqalign{ \sp(2.0)
ds^2 &= {}^{(4)}g_{\mu\nu} dx^{\mu} dx^{\nu},
\cr
     &= g_{ab} dx^a dx^b + \phi^2 ( d\theta^2 + \sin^2\theta d\varphi^2 ), 
\cr
\sp(3.0)} \eqno(4)$$
$$ \eqalign{ \sp(2.0)
D_{\theta} \Phi = A_{\theta} =  D_{\varphi} \Phi = A_{\varphi} = 0,
\cr
\sp(3.0)} \eqno(5)$$
where the two dimensional metric $g_{ab}$ and the radial function $\phi$ 
are the functions of only the two dimensional coordinates $x^a$.
After substituting (4) and (5) into (1) and then integrating over the 
angular coordinates $(\theta, \varphi)$, one obtains the following two 
dimensional effective action
$$ \eqalign { \sp(2.0)
S &= {1 \over 2} \int  d^2 x \sqrt{-g} \ \bigl[ 1 + g^{ab} \partial_a \phi 
\partial_b \phi +  {1 \over 2} R \phi^2 \bigr] 
\cr
&\qquad- \int  d^2 x \sqrt{-g} \ \phi^2 g^{ab} (D_a 
\Phi)^{\dag} D_b \Phi - {1 \over 4} \int  d^2 x \sqrt{-g} \ \phi^2 F_{ab} 
F^{ab}.
\cr
\sp(3.0)} \eqno(6)$$
 
The suitable ADM splitting of (1+1)-dimensional spacetime is given by 
$$ \eqalign{ \sp(2.0)
g_{ab} = \left(\matrix{ { - \alpha^2 + {\beta^2 \over \gamma}} & \beta \cr
              \beta & \gamma \cr} \right),
\cr
\sp(3.0)} \eqno(7)$$
and the normal unit vector $n^a$ orthogonal to the hypersurfaces 
$x^0 = const$ reads 
$$ \eqalign{ \sp(2.0)
n^a = ( {1 \over \alpha}, - {\beta \over {\alpha \gamma}}).
\cr
\sp(3.0)} \eqno(8)$$
In terms of this parametrization, the action (6) can be written as
$$ \eqalign{ \sp(2.0)
S &= \int d^2x L  =\int d^2x \ \bigl[ \ {1 \over 2} \alpha 
\sqrt{\gamma} \ \bigl\{ 1 - (n^a \partial_a 
 \phi)^2 + {1 \over \gamma}  (\phi^\prime)^2 - K n^a \partial_a (\phi^2) 
\cr
&\qquad+ {\alpha^\prime \over {\alpha\gamma}} \partial_1 (\phi^2) \bigr\} 
+ \alpha \sqrt{\gamma} \ \phi^2 \bigl\{{ | n^a D_a \Phi |^2 - {1 \over \gamma}
|D_1 \Phi |^2 }\bigr\} + {1 \over 2} \alpha \sqrt{\gamma} 
\ \phi^2 E^2 \bigr],
\cr
\sp(3.0)} \eqno(9)$$
where
$$ \eqalign{ \sp(2.0)
K = {1 \over \sqrt{-g}} \partial_a ( \sqrt{-g} n^a ) 
= {\dot \gamma \over {2\alpha\gamma}} - {\beta^\prime \over 
{\alpha\gamma}} + {\beta \over {2\alpha\gamma^2}} \gamma^\prime,
\cr
\sp(3.0)} \eqno(10)$$
$$ \eqalign{ \sp(2.0)
E&= {1 \over \sqrt{-g}} F_{01} = {1 \over \alpha \sqrt{\gamma}}    
(\dot A_1 - A_0^{\prime}),
\cr
\sp(3.0)} \eqno(11)$$
and ${\partial \over {\partial x^0}} = \partial_0$ and ${\partial \over 
{\partial x^1}} = \partial_1$ are also denoted by an overdot and a prime, 
respectively. 

Now the canonical conjugate momenta can be read off from the action (9). 
They are
$$ \eqalign{ \sp(2.0)
p_{\Phi} = \sqrt{\gamma} \ \phi^2 (n^a D_a\Phi)^{\dag},
\cr
\sp(3.0)} \eqno(12)$$
$$ \eqalign{ \sp(2.0)
p_{\Phi^{\dag}} = \sqrt{\gamma} \ \phi^2 (n^a D_a\Phi),
\cr
\sp(3.0)} \eqno(13)$$
$$ \eqalign{ \sp(2.0)
p_{\phi} = - \sqrt{\gamma} \ n^a \partial_a \phi - \sqrt{\gamma} \ K \phi,
\cr
\sp(3.0)} \eqno(14)$$
$$ \eqalign{ \sp(2.0)
p_{\gamma} = - {1 \over 4 \sqrt \gamma} n^a \partial_a (\phi^2),
\cr
\sp(3.0)} \eqno(15)$$
$$ \eqalign{ \sp(2.0)
p_A = \phi^2 E.
\cr
\sp(3.0)} \eqno(16)$$
Then the Hamiltonian can be calculated to be
$$ \eqalign{ \sp(2.0)
H &= \int dx^1 \ ( p_{\Phi} \dot \Phi + {p_{\Phi^{\dag}}} 
\dot \Phi^{\dag} 
+ p_{\phi} \dot \phi + p_{\gamma} \dot \gamma + p_A \dot A_1 - L ), 
\cr
  &= \int dx^1 \ ( \alpha H_0 + \beta H_1 + A_0 H_2 ), 
\cr
\sp(3.0)} \eqno(17)$$
where the constraints are explicitly given by
$$ \eqalign{ \sp(2.0)
H_0 &= {1 \over {\sqrt{\gamma} \phi^2}} p_{\Phi} {p_{\Phi^{\dag}}} - 
{\sqrt {\gamma} \over 2} - { (\phi^\prime)^2 \over {2 \sqrt{\gamma}}} 
+ \partial_1 ( {\partial_1 (\phi^2) \over {2 \sqrt{\gamma}}}) 
\cr
&\qquad+ {\phi^2 \over \sqrt{\gamma}} | D_1 \Phi |^2 - {2 \sqrt{\gamma} 
\over \phi} p_\phi p_\gamma + {2 \gamma \sqrt{\gamma} \over \phi^2} 
p_\gamma ^2  + {\sqrt{\gamma} \over {2 \phi^2}} p_A ^2 , 
\cr
\sp(3.0)} \eqno(18)$$
$$ \eqalign{ \sp(2.0)
H_1 = {1 \over \gamma} \ [p_\Phi D_1 \Phi + {p_{\Phi^{\dag}}} (D_1 \Phi)^{\dag}]
 + {1 \over \gamma} p_\phi \phi^\prime - 2  p_\gamma ^\prime - {1 \over 
\gamma} p_\gamma \gamma^\prime, 
\cr
\sp(3.0)} \eqno(19)$$
$$ \eqalign{ \sp(2.0)
H_2 = - ie \ (p_{\Phi} \Phi - p_{\Phi^{\dag}} \Phi^{\dag}) -  p_A ^\prime.
\cr
\sp(3.0)} \eqno(20)$$

Here for later convenience, let us derive equations of motion arising 
from the action (6):
$$ \eqalign{ \sp(2.0)
&{} - {2 \over \phi} \nabla_a \nabla_b \phi + {2 \over \phi} g_{ab} \nabla_c 
\nabla^c \phi + {1 \over \phi^2} g_{ab} \partial_c \phi \partial^c \phi 
- {1 \over \phi^2} g_{ab} 
\cr
&\qquad= - g_{ab} E^2 + 2 [ ( D_a \Phi)^{\dag}  D_b \Phi + ( D_b 
\Phi)^{\dag} D_a \Phi - g_{ab} ( D_c \Phi)^{\dag}  D^c \Phi ],
\cr
\sp(3.0)} \eqno(21)$$
$$ \eqalign{ \sp(2.0)
\nabla_a \nabla^a \phi - {1 \over 2} R \phi = \phi E^2 - 2 \phi ( D_a 
\Phi)^{\dag} D^a \Phi,
\cr
\sp(3.0)} \eqno(22)$$
$$ \eqalign{ \sp(2.0)
\nabla_a (\phi^2 E) = i e \phi^2 \sqrt{-g} \ \varepsilon_{ab} [ \Phi^{\dag} 
D^b \Phi - \Phi ( D^b \Phi)^{\dag} ],
\cr
\sp(3.0)} \eqno(23)$$
$$ \eqalign{ \sp(2.0)
D_a (\sqrt{-g} \ \phi^2 g^{ab} D_b \Phi) = 0,
\cr
\sp(3.0)} \eqno(24)$$
where $\varepsilon_{01} = - \varepsilon_{10} = +1$.

\vskip 32pt
\leftline{\bf 3. Canonical Formalism near Apparent Horizons}	
\par
Following the key observation mentioned in the introduction, we shall 
focus our thoughts on the canonical formalism in the vicinity of the 
horizons $\footnote 1 {P.Moniz has independently considered a similar 
model from a different motivation [19] (private communication). 
}$.  As a bonus of taking account of the spacetime region near the 
horizons, it turns  
out that the complicated constraints (18)-(20) can be cast into rather 
simple and technically tractable forms.

To fix the system of coordinates, let us introduce the 
 two dimensional coordinates $x^a$ by 
$$ \eqalign{ \sp(2.0)
x^a = (x^0, x^1) = (v - r, r),
\cr
\sp(3.0)} \eqno(25)$$
where the advanced time coordinate is defined as $v = t + r^*$ with the 
tortoise coordinate $dr^* = {dr \over {-g_{00}}}$. Since we wish to 
consider the Reissner-Nordstrom charged black hole,  we shall fix the 
gauge freedoms corresponding to the two dimensional 
reparametrization invariances by the gauge conditions
$$ \eqalign{ \sp(2.0)
g_{ab} &= \left(\matrix{ { - \alpha^2 + {\beta^2 \over \gamma}} & \beta \cr
              \beta & \gamma \cr} \right),
\cr
 &= \left(\matrix{ -(1 - {2M \over r} + {Q^2 \over r^2})   &  {2M \over 
 r} - {Q^2 \over r^2} \cr
              {2M \over r} - {Q^2 \over r^2} & 1 + {2M \over r} - {Q^2 
              \over r^2} } \right),
\cr
\sp(3.0)} \eqno(26)$$
where the black hole mass $M$ and the electric charge $Q$ are now in 
general the  
function of the two dimensional coordinates $x^a$. Notice that we have 
not fixed the gauge symmetries completely because we wish to consider a 
dynamical black hole whose mass and charge change in time under an 
influence of motion of charged matter field. However, the remaining 
gauge freedoms are effectively fixed in making the assumptions of 
dynamical fields near the horizons later. Moreover, the $U(1)$ gauge 
symmetries are fixed by the gauge conditions 
$$ \eqalign{ \sp(2.0)
A_0 = A_1 = {Q \over r}.
\cr
\sp(3.0)} \eqno(27)$$

{}From (25) and (26) the two dimensional line element takes a form
$$ \eqalign{ \sp(2.0)
ds^2 &= g_{ab} dx^a dx^b,
\cr
     &= -(1 - {2M \over r} + {Q^2 \over r^2}) dv^2 + 2 dv dr.
\cr
\sp(3.0)} \eqno(28)$$
This is precisely the charged Vaidya metric for the Reissner-Nordstrom 
black hole. For a dynamical black hole, it is convenient to introduce the 
local horizon, i.e., the apparent horizon, instead of the  
event horizon since a genuine event horizon is a complicated global 
object.  The apparent horizons are now defined as
$$ \eqalign{ \sp(2.0)
r_{\pm} = M \pm \sqrt{M^2 - Q^2},
\cr
\sp(3.0)} \eqno(29)$$
where $r_{+}$ and $r_{-}$ denote respectively the outer apparent horizon 
and the inner Cauchy horizon. We will assume that $M > Q$ as an initial 
condition. As a familiar fact, in the extremal limit $M = Q$ these two 
horizons coincide, and for $M < Q$ they are absent and a naked 
singularity is visible to external observers.

Near the apparent horizons (29), (26) yields
$$ \eqalign{ \sp(2.0)
\alpha \approx {1 \over \sqrt{2}} , \ \beta \approx 1, \ \gamma = {1 
\over \alpha^2} \approx 2,
\cr
\sp(3.0)} \eqno(30)$$
From now on we shall use $\approx$ to indicate the equalities which hold 
approximately near the apparent horizons. Note that  
the dynamical degrees of freedom representing ``graviton'' $\gamma$ are 
effectively fixed in (30). Also note that we are concerned with the 
spacetime regions where $x^0$ plays a role of time, so $x^1 = r$ 
coordinate must take the ranges of $0 < r < r_{-}$ and $r_{+} < r < 
\infty$.  An analogous formulation which holds in the range of $r_{-} < r 
< r_{+}$ can be easily built in a similar way to the reference [13]. 
Now let us make physically plausible assumptions near the apparent 
horizons  
$$ \eqalign{ \sp(2.0)
\Phi \approx \Phi(v), \ M \approx M(v), \ Q \approx Q(v), \ \phi \approx r. 
\cr
\sp(3.0)} \eqno(31)$$
A consistency of the assumptions (31) with the field equations 
(21)-(24) will be 
discussed in the end of this section. Given the assumptions (31), 
the canonical conjugate momenta (12)-(16) become
$$ \eqalign{ \sp(2.0)
p_{\Phi} &\approx  \phi^2 \ (\partial_v \Phi^{\dag} - i e A_0 \Phi^{\dag}),
\cr
p_{\Phi^{\dag}} &\approx  \phi^2 \ (\partial_v \Phi + i e A_0 \Phi),
\cr
p_{\phi} &\approx 1 + {1 \over 2} (\partial_v M - {Q \over r} \partial_v 
Q) + {3 \over 2} (- {M \over r} + {Q^2 \over r^2}),
\cr
p_{\gamma} &\approx {1 \over 4} \phi,
\cr
p_A &\approx Q.
\cr
\sp(3.0)} \eqno(32)$$
It is remarkable that the Hamiltonian constraint $H_0 = 0$ becomes 
proportional to the supermomentum constraint $H_1 = 0$ in the vicinity of 
the apparent horizons 
$$ \eqalign{ \sp(2.0)
\sqrt{2} H_0 &\approx 2  H_1,
\cr
             &\approx {2 \over \phi^2} p_{\Phi} p_{\Phi^{\dag}} - 2 
             p_\phi - 2 + {5M \over \phi} - {Q^2 \over \phi^2},
\cr
\sp(3.0)} \eqno(33)$$
since the time translation is frozen at the apparent horizons in this 
coordinate system [12]. In addition, along with $p_A \approx Q$ in (32) 
the $U(1)$ constraint $H_2 = 0$ produces the equation 
$$ \eqalign{ \sp(2.0)
\partial_v Q \approx - i e \ (\Phi p_{\Phi} - \Phi^{\dag} 
p_{\Phi^{\dag}}).   
\cr
\sp(3.0)} \eqno(34)$$
This will be used when evaluating the expectation value of $\partial_v Q$ 
in section 5. In the formalism developed so far, the dynamical degrees of 
freedom are contained entirely in the matter field, and the other fields 
are determined by solving the constraint equations by help of the gauge 
conditions and the assumptions (31). This reduction of the dynamical 
degrees of freedom comes from the gauge symmetries and the assumptions 
adopted near the apparent horizons.

Before closing this section, we would like to discuss the consistency of 
the assumptions (31) with equations of motion (21)-(24). After a tedious 
but straightforward calculation, if we assume (31) to hold true except 
$\Phi$,  
equations of motion (21)-(24) lead to the following equations: 
$$ \eqalign{ \sp(2.0)
\partial_r \Phi \approx 0,
\cr
\sp(3.0)} \eqno(35)$$
$$ \eqalign{ \sp(2.0)
(D_v \Phi)^{\dag} D_v \Phi \approx {1 \over {2 r}} ({\partial_v M \over r} 
- {{Q \partial_v Q} \over r^2}),
\cr
\sp(3.0)} \eqno(36)$$
$$ \eqalign{ \sp(2.0)
\partial_v Q \approx  i e \phi^2 \ [ \Phi^{\dag} D_v \Phi - \Phi (D_v 
\Phi)^{\dag} ],   
\cr
\sp(3.0)} \eqno(37)$$
$$ \eqalign{ \sp(2.0)
2 r^2 \partial_r \partial_v \Phi + 2 r \partial_v \Phi + i e Q \Phi \approx 
0 ,
\cr
\sp(3.0)} \eqno(38)$$
$$ \eqalign{ \sp(2.0)
\partial_r \partial_v \Phi \approx \partial_v \partial_r \Phi.
\cr
\sp(3.0)} \eqno(39)$$
As shown in the reference [11], at least in the case of the Schwarzschild black 
hole and the neutral scalar matter field, the consistency is explicitly 
proved by finding the solution
$$ \eqalign{ \sp(2.0)
\Phi(v) \approx  \int^v dv {1 \over {2 M(v)}} \sqrt{\partial_v M \over 
2},
\cr
\sp(3.0)} \eqno(40)$$
which implies that the increase of the black hole mass, $\partial_v M > 
0$, is classically allowed, but the loss of the black hole mass, 
$\partial_v M < 0$, that is , the Hawking radiation, is classically 
forbidden and occurs only through the quantum-mechanical tunneling 
effect. Unfortunately, in the Reissner-Nordstrom geometry with the 
charged scalar matter field it seems to difficult to find such an analytic 
solution owing to the highly nonlinear structure of (36) and (37).  
However, we believe that our assumptions (31) are not in unjustified leap 
in the dark from physical viewpoint 
and that the physically plausible results which will be obtained 
in later sections would support the assumptions.

\vskip 32pt
\leftline{\bf 4. Hawking Radiation}	
\par
In the previous section, we have constructed the canonical formalism 
holding near the apparent horizons. We are now ready to perform the 
canonical quantization of it. After completing this, as its concrete 
applicaton we will  
investigate the Hawking radiation by purely quantum-mechanical 
treatment. 

For simplicity, in this section for simplicity let us consider the case 
of the neutral matter field $\Phi^{\dag} = 
\Phi$, for which the constraint $H_2 = 0$ makes no sense since now the 
black hole charge $Q$ takes a fixed constant. By imposing the 
constraint (33) as an operator equation one obtains the Wheeler-DeWitt 
equation $\footnote 1 {We have changed the coefficient in front of the 
matter action from ${1 \over 4 \pi}$ to ${1 \over 8 \pi}$ in (1) in the 
case of the neutral scalar field, which 
gives rise to the modification from ${2 \over \phi^2} p_{\Phi} 
p_{\Phi^{\dag}}$ to ${1 \over \phi^2} p_{\Phi} ^2$ in (33).}$ 
$$ \eqalign{ \sp(2.0)
( - {1 \over \phi^2} {\partial^2 \over {\partial \Phi^2}} + 2 i {\partial 
\over {\partial \phi}} + {M \over \phi} + {Q^2 \over \phi^2} ) \Psi = 
0.  
\cr
\sp(3.0)} \eqno(41)$$
The simplest way to find a special solution of this Wheeler-DeWitt 
equation is to use the method of separation of variables. The result reads 
$$ \eqalign{ \sp(2.0)
\Psi = (  B e^{-\sqrt A \Phi} + C e^{\sqrt A \Phi}) e^{ {i \over 4} ( \phi 
+ {2 A - 3 Q^2 \over \phi})},
\cr
\sp(3.0)} \eqno(42)$$
where $A$, $B$, and $C$ are integration constants. Without losing a 
generality, we shall choose the boundary condition $B = 0$. Here if one 
defines an expectation value $< \cal O >$ of an operator $\cal O$ as 
$\footnote 2 {Here we have explicitly written the definition in the case 
of the complex scalar field for later convenience. Specification to the 
case of the real scalar field is obvious.}$ 
$$ \eqalign{ \sp(2.0)
< {\cal O} > = {1 \over {\int d\Phi^{\dag} d\Phi e^{-|\Phi|^2} |\Psi|^2 
}}  \int d\Phi^{\dag} d\Phi e^{-|\Phi|^2} \Psi^{\dag} {\cal O} 
\Psi,
\cr
\sp(3.0)} \eqno(43)$$
then $< \partial_v M >$ can be calculated to be
$$ \eqalign{ \sp(2.0)
< \partial_v M > = - {A \over {< r_{\pm} ^2 >}}.
\cr
\sp(3.0)} \eqno(44)$$
It is of interest to notice that different choices of the integration 
constant $A$ lead to different physical pictures. Namely, a negative 
choice $A = - k_1 ^2$ gives us a physical picture of the absorption of 
the external matters by a black hole 
$$ \eqalign{ \sp(2.0)
< \partial_v M > =  {k_1 ^2 \over { < r_{\pm} ^2 > }},
\cr
\sp(3.0)} \eqno(45)$$
with the wave function having a form of the scalar wave propagating into 
black hole in the superspace 
$$ \eqalign{ \sp(2.0)
\Psi = C \ e^{-i | k_1 | \Phi(v) + {i \over 4} (\phi - {{2 k_1 ^2 + 3 
Q^2}  \over \phi})}.
\cr
\sp(3.0)} \eqno(46)$$
On the other hand, a positive constant $A = k_2 ^2$ yields the Hawking 
radiation  
$$ \eqalign{ \sp(2.0)
< \partial_v M > = - {k_2 ^2 \over { < r_{\pm} ^2 > }},
\cr
\sp(3.0)} \eqno(47)$$
for which this time the 
physical state has an exponentially 
damping-like form in the classically forbidden region showing the quantum 
tunneling 
$$ \eqalign{ \sp(2.0)
\Psi = C \ e^{- | k_2 | \Phi(v) + {i \over 4} (\phi + {{2 k_2 ^2 - 3 
Q^2}  \over \phi})}.
\cr
\sp(3.0)} \eqno(48)$$
Moreover, if one assumes that the black hole charge is zero, the result 
for the outer apparent horizon exactly equals to the result calculated 
by Hawking in the semiclassical approach [1]
$$ \eqalign{ \sp(2.0)
< \partial_v M > = - {k_2 ^2 \over { 4 < M^2 > }}.
\cr
\sp(3.0)} \eqno(49)$$
It is worth pointing out the difference 
between the Hawking formulation and the present one. In the Hawking's 
semiclassical approach the gravitational field is fixed as a classical 
background and only the matter field is treated to be quantum-mechanical. 
By contrast, our present formulation is purely quantum-mechanical in that 
both the gravitational field and the matter field are considered to be 
quantum fields.

\vskip 32pt
\leftline{\bf 5. Cosmic Censorship in Quantum Gravity}	
\par
Being encouraged with success of deriving a physically interesting 
results with respect to the Hawking radiation by the quantum 
gravity in the previous section, 
we   
now turn our attention to the problem of cosmic censorship. To begin 
with, in order to clarify the contents of the problem which we wish to 
solve, let us remind you of the physical setting of the problem. 

At the outset, suppose that there is a stable Reissner-Nordstrom 
electrically charged black hole satisfying the relation $Q < M$ which 
means that this black hole has two event horizons at $r = r_{\pm} = M \pm 
\sqrt{M^2 - Q^2}$. Then we would increase the black hole charge $Q$ 
faster than its mass $M$ by sending a lightlike flux of charged scalar 
matters with $q > m$ to this black hole from the exterior. Here $q$ and 
$m$ denote respectively the charge and the energy of these scalar 
matters.  As a result, if the black hole  
would become to have the charge larger than its mass, i.e., $Q > M$, 
cosmic censorship would be violated in this process since for $Q > M$ 
there is no more horizon so that a naked singularity is visible to 
external observers. Roughly speaking, this would be a test that one 
tries to destroy the outer horizon of a black hole by forcing it to meet 
the inner horizon. As mentioned in section 1, it was proved that cosmic 
censorship could never be violated this way in classical theory of 
general relativity [9]. Our aim in this section is to examine whether 
quantum gravity would modify this classical result or not.

Since we have specified the contents of the problem, we challenge to 
solve the  
problem by applying the quantum gravity developed in previous sections. 
Since the charged matter field plays an essential role here, one has to 
solve the Wheeler-DeWitt equation arising from (33) which is given by
$$ \eqalign{ \sp(2.0)
( - {2 \over \phi^2} {\partial^2 \over {\partial \Phi \partial 
\Phi^{\dag}}} + 2 i {\partial \over {\partial \phi}} - 2 + {5 M \over 
\phi} - {Q^2 \over \phi^2} ) \Psi = 0.  
\cr
\sp(3.0)} \eqno(50)$$
Again a use of the method of separation of variables yields a 
special solution  
$$ \eqalign{ \sp(2.0)
\Psi =   C \ e^{A \Phi^{\dag} + B \Phi +  i  ( {{A B + {1 \over 2} Q^2} 
\over \phi} - \phi + {5 \over 2} M \log{\phi} ) }.
\cr
\sp(3.0)} \eqno(51)$$
As a simple check, it is valuable to calculate $< p_A >$ by 
inserting the physical state (51) into the definition of an expectation 
value (43) . 
The result is $< p_A > = Q$, which is consistent with the last equation 
in (32).

Next let us evaluate the expectation value of the change rate of 
charge $< \partial_v Q >$. 
After a simple calculation, we obtain $\footnote 1 {Even if we 
introduce the weight factor $e^{- K |\Phi|^2}$ instead of $e^{- 
|\Phi|^2}$ in the definition of an expectation value (43), the 
coefficient $K$ can  be  
``renormalized'' into the electric charge by $e_R = {e \over K}$. }$ 
$$ \eqalign{ \sp(2.0)
< \partial_v Q > = e \ ( |A|^2 - |B|^2 ).
\cr
\sp(3.0)} \eqno(52)$$
First of all, this equation indicates that if $\Phi^{\dag}$ creates the 
particle with a  
charge $e$, $\Phi$ creates the particle with the opposite charge $- e$ in 
a language of the field theory. The more important point is that the 
right-hand side of (52) is a certain constant so the black hole charge 
$Q$ increases or decreases at the constant rate according to the motion 
of external charged matters across the apparent horizons. At this point, 
we assume that $e \gg m$ and $|A| > |B|$, which corresponds to the 
physical setting described above that charged matters with a large $q 
\over m$ ratio flow in a black hole to urge the black hole 
charge rather than its mass to increase.

In addition, from (32) or (33), we obtain the equations
$$ \eqalign{ \sp(2.0)
< \partial_v r_{\pm} > = - {4 A B \over { < r_{\pm} ^2 > }}.
\cr
\sp(3.0)} \eqno(53)$$
Then adding or subtracting these equations each other gives
$$ \eqalign{ \sp(2.0)
< \partial_v M > = - 4 A B \ {{2 M^2 - Q^2} \over Q^4},
\cr
\sp(3.0)} \eqno(54)$$
$$ \eqalign{ \sp(2.0)
< \partial_v ( r_{+} - r_{-} ) > = 16 A B \ {{M \sqrt{M^2 - Q^2}} \over 
Q^4},
\cr
\sp(3.0)} \eqno(55)$$
where $r_{\pm} = M \pm \sqrt{M^2 - Q^2}$ is used. As told in the above, 
since we have in mind a physical situation where external charged matters 
come in a black hole across the apparent horizons, the black hole mass 
must increase as time passes by. This demands us to choose the product of 
integration  
constants $AB$ to be a negative constant, e.g., $- {k_1 ^2 \over 16}$ in 
(54). Then (55) becomes  
$$ \eqalign{ \sp(2.0)
< \partial_v ( r_{+} - r_{-} ) > = - k_1 ^2 \ {{M \sqrt{M^2 - Q^2}} \over 
Q^4}.
\cr
\sp(3.0)} \eqno(56)$$
Eq.(56) gives very important informations about the behavior of 
the outer apparent horizon and the inner Cauchy horizon. Namely, as 
external  
matters are accumulated into a black hole, the inner Cauchy horizon 
approaches the outer apparent horizon but the relative approaching speed 
of the horizons, which is quantitatively controlled by the right-hand 
side of this  
equation, descreases gradually.  Once the extremal limit $Q = M$ is 
realized, its relative speed vanishes so that both horizons coalesce, and 
afterward expand together. The reason why the expansion of the coalesced 
horizons does not stop is that in this model external charged matters 
constantly flow into a black hole as understood from (52).  Note that the 
above picture is consistent with (54) where even in the extremal limit 
the increase rate of the black hole mass is positive. One of the 
most important points here is that the inner Cauchy horizon never goes beyond 
the outer apparent horizon, which means that cosmic censorship remains 
true also in quantum gravity as in classical gravity.

\vskip 32pt
\leftline{\bf 6. Conclusion}	
\par
In this article, we have investigated weak cosmic censorship from the 
viewpoint of quantum gravity, and reached the result that cosmic 
censorship is not violated as in classical theory of general 
relativity. Although we have obtained the same conclusion as the 
classical gravity, the intermediate way of calculations are quite 
different between them. We should clarify their relationship in the 
future. It is true that cosmic censorship cannot be proved 
in general by studying a specific model, but we believe that the physical 
setting that we have examined is so fundamental that the results obtained 
here would further the more advanced study. 

In previous works [11-14], we have so far applied the quantum gravity 
holding near the horizons and/or the singularities in black holes to 
several problems of quantum black holes, by which we have gained the 
physically interesting results. Since the modern unsolved problems 
associated with quantum black holes such as the black hole entropy and 
the fate of the endpoint of black hole evaporation e.t.c. are all related 
to the properties of the horizons and the singularities, it seems to be 
natural that our formalism has had a wide applications toward these 
problems. In other words, 
there should be a hopeful possibility that the quantum physics in the 
vicinity of the horizons and the singularities determine the physical 
properties of quantum black holes essentially. Incidentally, the idea 
that the (stretched) horizon would play an important role in 
understanding the phenomena about black holes was previously 
advocated as the Membrane Paradigm [20-22]. The results of this paper might 
partly support this conjecture.

\vskip 32pt
\leftline{\bf References}
\centerline{ } %
\par
\item{[1]} S.W.Hawking, Comm. Math. Phys. {\bf 43} (1975) 199.

\item{[2]} D.Page, in Proceedings of the Fifth Canadian Conference on 
General Relativity and Relativistic Astrophysics, edited by 
R.B.Mann et al. (World Scientific, Singapore, 1994). 

\item{[3]} J.D.Bekenstein, Phys. Rev. {\bf D7} (1973) 2333.

\item{[4]} R.Penrose, Rev. del Nuovo Cimento {\bf 1} (1969) 252.

\item{[5]} A.Hosoya, Class. Quant. Grav. {\bf 12} (1995) 2967.

\item{[6]} A.Hosoya and I.Oda, ``Black Hole Singularity and Generalized 
Quantum Affine Parameter'', TIT/HEP-334/COSMO-73, EDO-EP-5, gr-qc/9605069.

\item{[7]} R.Penrose, in Theoretical Principles in Astrophysics and 
General Relativity, edited by N.R.Lebovitz et al. (University of Chicago 
Press, Chicago, 1978); W.Israel, Can. J. Phys. {\bf 64} (1986) 120.

\item{[8]} S.W.Hawking, ``Nature of Space and Time'', hep-th/9409195.

\item{[9]} R.Wald, Ann. Phys. {\bf 82} (1974) 548.

\item{[10]} A.Tomimatsu, Phys. Lett. {\bf B289} (1992) 283.

\item{[11]} A.Hosoya and I.Oda, Prog. Theor. Phys. {\bf 97} (1997) 233.

\item{[12]} I.Oda, ``Evaporation of Three Dimensional Black Hole in 
Quantum Gravity'', EDO-EP-9, gr-qc/9703055.

\item{[13]} I.Oda, ``Mass Inflation in Quantum Gravity'', EDO-EP-8, 
gr-qc/9701058.

\item{[14]} I.Oda, ``Quantum Instability of Black Hole Singularity 
in Three Dimensions'', EDO-EP-10, gr-qc/9703056.

\item{[15]} P.Hajicek, Phys. Rev. {\bf D30} (1984) 1178; P.Thomi, B.Isaak, 
and P.Hajicek, Phys.  Rev. {\bf D30} (1984) 1168.

\item{[16]} P.A.M.Dirac, Lectures on Quantum Mechanics (Yeshiva University, 
1964). 

\item{[17]} P.Arnowitt, S.Deser, and C.W.Misner, in Gravitation: An 
Introduction to Current Research, edited by L.Witten (Wiley, New York, 
1962).

\item{[18]} C.W.Misner, K.S.Thorne, and J.A.Wheeler, Gravitation (Freeman, 
1973). 

\item{[19]} P.Moniz, preprint to appear.

\item{[20]} K.S.Thorne, R.H.Price and D.A.Macdonald, Black Hole: The 
Membrane Paradigm (Yale University Press, 1986).

\item{[21]} G.'t Hooft, Nucl. Phys. {\bf B335} (1990) 138; Phys. Scripta 
{\bf T15} (1987) 143; ibid. {\bf T36} (1991) 247.

\item{[22]} I.Oda, Int. J. Mod. Phys. {\bf D1} (1992) 355; Phys. Lett. 
{\bf B338} (1994) 165; Mod. Phys. Lett. {\bf A10} (1995) 2775.

\endpage
%

%
\bye